\documentclass[12pt]{article}
\if@twoside\oddsidemargin25mm
\else\oddsidemargin25mm\evensidemargin25mm\marginparwidth25mm\fi%
\begin{document}
\newcommand{\dr}{\raise.3ex\hbox{$\stackrel{\leftarrow}{\partial }$}{}}
\newcommand{\dl}{\raise.3ex\hbox{$\stackrel{\rightarrow}{\partial}$}{}}
\newcommand{\eqn}[1]{(\ref{#1})}
\newcommand{\ft}[2]{{\textstyle\frac{#1}{#2}}}
\newcommand{\dkt}{\delta _{KT}}
\newcommand{\QED}{{\hspace*{\fill}\rule{2mm}{2mm}\linebreak}}
\renewcommand{\theequation}{\thesection.\arabic{equation}}
\csname @addtoreset\endcsname{equation}{section}
\newsavebox{\uuunit}
\sbox{\uuunit}
    {\setlength{\unitlength}{0.825em}
     \begin{picture}(0.6,0.7)
        \thinlines
        \put(0,0){\line(1,0){0.5}}
        \put(0.15,0){\line(0,1){0.7}}
        \put(0.35,0){\line(0,1){0.8}}
       \multiput(0.3,0.8)(-0.04,-0.02){12}{\rule{0.5pt}{0.5pt}}
     \end {picture}}
\newcommand {\unity}{\mathord{\!\usebox{\uuunit}}}
\newcommand  {\Rbar} {{\mbox{\rm$\mbox{I}\!\mbox{R}$}}}
\newcommand  {\Hbar} {{\mbox{\rm$\mbox{I}\!\mbox{H}$}}}
\newcommand {\Cbar}
    {\mathord{\setlength{\unitlength}{1em}
     \begin{picture}(0.6,0.7)(-0.1,0)
        \put(-0.1,0){\rm C}
        \thicklines
        \put(0.2,0.05){\line(0,1){0.55}}
     \end {picture}}}
\newsavebox{\zzzbar}
\sbox{\zzzbar}
  {\setlength{\unitlength}{0.9em}
  \begin{picture}(0.6,0.7)
  \thinlines
  \put(0,0){\line(1,0){0.6}}
  \put(0,0.75){\line(1,0){0.575}}
  \multiput(0,0)(0.0125,0.025){30}{\rule{0.3pt}{0.3pt}}
  \multiput(0.2,0)(0.0125,0.025){30}{\rule{0.3pt}{0.3pt}}
  \put(0,0.75){\line(0,-1){0.15}}
  \put(0.015,0.75){\line(0,-1){0.1}}
  \put(0.03,0.75){\line(0,-1){0.075}}
  \put(0.045,0.75){\line(0,-1){0.05}}
  \put(0.05,0.75){\line(0,-1){0.025}}
  \put(0.6,0){\line(0,1){0.15}}
  \put(0.585,0){\line(0,1){0.1}}
  \put(0.57,0){\line(0,1){0.075}}
  \put(0.555,0){\line(0,1){0.05}}
  \put(0.55,0){\line(0,1){0.025}}
  \end{picture}}
\newcommand{\Zbar}{\mathord{\!{\usebox{\zzzbar}}}}
\newcommand{\cmap}{{\bf c} map}
\newcommand{\rmap}{{\bf r} map}
\newcommand{\crmap}{{{\bf c}$\scriptstyle\circ${\bf r} map}}
\newcommand{\Ka}{K\"ahler}
\newcommand{\qu}{quaternionic}
\newcommand{\Al}{Alekseevski\v{\i}}
\def\ib{{\bar \imath}}
\def\jb{{\bar \jmath}}
\renewcommand{\sp}{{Sp\left( 2n+2,\Rbar\right)}}
\renewcommand{\a}{\alpha}
\renewcommand{\b}{\beta}
\renewcommand{\c}{\gamma}
\renewcommand{\d}{\delta}
\newcommand{\pa}{\partial}
\newcommand{\g}{\gamma}
\newcommand{\G}{\Gamma}
\newcommand{\A}{\Alpha}
\newcommand{\B}{\Beta}
\newcommand{\D}{\Delta}
\newcommand{\e}{\epsilon}
\newcommand{\E}{\Epsilon}
\newcommand{\z}{\zeta}
\newcommand{\Z}{\Zeta}
\newcommand{\K}{\Kappa}
\renewcommand{\l}{\lambda}
\renewcommand{\L}{\Lambda}
\newcommand{\m}{\mu}
\newcommand{\M}{\Mu}
\newcommand{\n}{\nu}
\newcommand{\N}{\Nu}
\newcommand{\x}{\chi}
\newcommand{\X}{\Chi}
\newcommand{\p}{\pi}
\newcommand{\R}{\Rho}
\newcommand{\s}{\sigma}
\renewcommand{\S}{\Sigma}
\renewcommand{\t}{\tau}
\newcommand{\T}{\Tau}
\newcommand{\y}{\upsilon}
\newcommand{\Y}{\upsilon}
\renewcommand{\o}{\omega}
\newcommand{\q}{\theta}
\newcommand{\h}{\eta}


\begin{titlepage}
\begin{flushright} THU-95/13\\ KUL-TF-95/13\\ hep-th/9505097
\end{flushright}
\vfill
\begin{center}
{\large\bf Isometries of special manifolds${}^\dagger$ }   \\
\vskip 7.mm
{B. de Wit }\\
\vskip 0.1cm
{\em Institute for Theoretical Physics} \\
{\em Utrecht University}\\
{\em Princetonplein 5, 3508 TA Utrecht, The Netherlands} \\[5mm]
 A. Van Proeyen${}^*$\\
\vskip 1mm
{\em Instituut voor theoretische fysica}\\
{\em Universiteit Leuven, B-3001 Leuven, Belgium}
\end{center}
\vfill

\begin{center}
{\bf ABSTRACT}
\end{center}
\begin{quote}
We describe special \Ka\ geometry, special \qu\ manifolds, and very
special real manifolds and analyze the structure of their
isometries.  The
classification of the homogeneous manifolds of these types is
presented.
\vfill      \hrule width 5.cm
\vskip 2.mm
{\small\small
\noindent $^\dagger$ Based on invited talks given at the Meeting on
Quaternionic Structures in Mathematics and Physics, Trieste,
September 1994; to be published in the proceedings.}\\
${}^*$ Onderzoeksleider, N.F.W.O., Belgium
\end{quote}
\begin{flushleft}
May 1995
\end{flushleft}
\end{titlepage}

\section{Introduction}

Quaternionic spaces with a transitive solvable group of motions
have been classified by \Al\ twenty years ago \cite{Aleks}.
More recently it was discovered that these so-called {\em normal}
quaternionic spaces and their classification are related to supergravity
coupled to abelian gauge fields \cite{CecFerGir,Cecotti,ssss}.
\par
Supergravity theories are invariant under local (i.e. space-time
dependent) supersymmetry transformations. Under such
transformations bosonic (commuting) and fermionic
(anticommuting) fields transform into each
other with parameters that are themselves anticommuting Lorentz
spinors. Extended supersymmetry implies that we are dealing with
$N$ independent supersymmetry transformations, each described
by a separate anticommuting spinorial parameter. The number of
supersymmetry generators (`supercharges') is thus equal to $N$
times the dimension of the (smallest) spinor representation. For
realistic supergravity this number of supercharge components cannot
exceed 32. As 32 is the number of components of a Lorentz
spinor in $d=11$ space-time dimensions, it follows that realistic
supergravity theories can only exist for dimensions $d\leq 11$. A
characteristic feature of the algebra containing the supersymmetry
generators is that it contains the space-time translation
operators. Therefore local supersymmetry leads to space-time
diffeomorphisms and the invariant actions are generalizations of
the Einstein-Hilbert action.
    \par
By its very nature supersymmetry implies the presence of both bosonic and
fermionic fields carrying integer and half-integer spin,
respectively. The
graviton, the particle described by the
space-time metric, has spin 2. The `gravitino', the particle
associated with the fermionic gauge field of supersymmetry, has
spin 3/2. Massless spin-3/2 particles are described by the
Rarita-Schwinger action. Obviously the number of gravitini must
be equal to $N$. All other particles carry spin less than 3/2. In a
four-dimensional space-time they are usually described by vector,
spinor and spinless fields and in the action they appear in
generalizations of the Yang-Mills, Dirac and Klein-Gordon
actions.
\par
The type of supergravity is thus characterised by the numbers $d$
and $N$. For instance, for the physical
$d=4$ dimensional space-time, one can have
supergravity theories with $1\leq N\leq 8$.  For any $N$ and $d$
there is a {\em pure} supergravity theory, having physical states
with spins ranging from 0 to 2.  If the  number of
supercharge components does not exceed 16, one can have couplings
with (supersymmetric) matter,
which has spin $s\leq 1$. For the purpose of our work, we shall be
dealing with 8 supercharge components, corresponding to $N=4$ in three,
$N=2$ in four and five, and $N=1$ in six space-time dimensions.
\par
Of particular interest for geometry are the spinless fields, denoted by
$\phi^i(x)$, which define a map from the $d$-dimensional Minkowskian
space-time, with coordinates $x^\mu$ with $\mu=1,..., d$, to some
`target space'.
In the physics literature, such a model is called a nonlinear
sigma model. The kinetic terms in the action read
\begin{equation}
S=-\ft12\int d^dx \;\sqrt{\det g(x)}\, g^{\mu\nu}(x)\,
\frac{\partial\phi^i(x)}{\partial x^\mu}  \,
\frac{\partial\phi^j(x)}{\partial x^\nu} \, G_{ij}(\phi(x))\ ,
\end{equation}
where $g^{\mu\nu}(x)$ refers to the space-time
metric. $G_{ij}(\phi)$ defines the target-space
metric corresponding to the invariant line element
\begin{equation}
{\rm d}s^2= G_{ij}(\phi)\, {\rm d}\phi^i {\rm d}\phi^j\ .
\end{equation}
\par
A crucial point is that supersymmetry severely restricts the
possible target-space geometries. As clearly exhibited in
table~\ref{tbl:mandN},
the more supercharge components one has,
the more restrictions one finds. For large $N$ and/or $d$ the
target-space manifold is a unique symmetric space. When the
\begin{table}[ht]
\caption{Restrictions on target-space manifolds
according to the type of supergravity theory. The rows are arranged
such that the number $\kappa$ of supercharge components is constant.
${\cal M}$ refers to a general Riemannian
manifold, $SK$ to `special \Ka', $VSR$ to
`very special real' and $Q$ to \qu\ manifolds. }
\label{tbl:mandN}
\begin{center}\begin{tabular}{ccccccc}\hline
$\kappa$&  d=3 & d=4 & d=5 & d=6& & \\ \hline
2  & $N=1$&&&&& \\
{}~&${\cal M}$ &&&&&  \\[1mm]   \hline
4&  $N=2$  & $N=1$ & &  &  & \\
{}~  &  \Ka  & \Ka   & &    &  &  \\[1mm] \hline
6&$N=3$ &&&&&\\
{}~&$Q$& &&&&\\[1mm]\hline
8&  $N=4$ & $N=2$        & $N=2$    & $N=1$  & &   \\
{}~&   $Q\oplus Q$   &  $SK\oplus Q$ & $VSR\oplus Q$ & $\O \oplus Q
$ & & \\[1mm]
\hline
10&$N=5$ &&&&&\\
{}~&${Sp(2,n)\over Sp(2)\otimes Sp(n)}$&& &&&\\[1mm]\hline
12&$N=6$ &&&&&\\
{}~&${SU(4,n)\over S(U(4)\otimes U(n))}$& &&&\\[1mm]\hline
16   & N=8 & $N=4$  & ... &... & $\rightarrow $& $N=1$      \\
{}~  & ${SO(8,n)\over SO(8)\otimes SO(n)}$
    & $\frac{SU(1,1)}{U(1)}\otimes \frac{SO(6,n)}{SO(6)\otimes SO(n)}$  & ... &
.. & $\to$&
$d=10$\\[1mm] \hline
18  & $N=9$ & ~&~ & ~ && \\
{}~  & ${F_{4(-20)}\over SO(9)}$ & ~  & ~ & ~&&\\[1mm] \hline
20 & N=10 & $N=5$  & ... & ~ & &   \\
{}~  & ${E_{6(-14)}\over SO(10)\otimes SO(2)}$ &
    $\frac{SU(5,1)}{U(5)}$  & ... & ~ &&\\[1mm] \hline
24 & N=12 & $N=6$  & $N=6$ & ... &    &\\
{}~  & ${E_{7(-5)}\over SO(12)\otimes SO(3)}$ &
$\frac{SO^\ast(12)}{U(6)}$  & ${SU^\ast(6)\over USp(6)}$ &...&  &
\\[1mm] \hline
32 & N=16 & $N=8$  & $N=8$&...  & $\rightarrow $ &$N=1$ \\
{}~  & ${E_{8(+8)}\over SO(16)}$ & $\frac{E_{7(+7)}}{SU(8)}$  &
${E_{6(+6)}\over USp(8)}$ &... &$\to$ & $d=11$ \\[1mm] \hline
\end{tabular}\end{center}
\end{table}
number of supercharge components equals 16, the target-space
geometry is fixed once the number of matter multiplets is
given (for $d=4$ each of them must
contain a spin-1 field). This row continues to $N=1$, $d=10$.
Beyond 16 supercharge components there
is no freedom left. The row with 32 supercharge components
continues to $N=1$, $d=11$. On the other hand in
$d=3$, $N=1$ any Riemannian manifold can occur. For $N=2$
in $d=3$, and $N=1$ in $d=4$, supersymmetry induces
a natural complex structure, and the manifold is \Ka ian. For
higher $N$ one finds quaternionic manifolds; the three complex
structures are again closely related to the supersymmetry
transformations. For the
purpose of this paper the row with $\kappa=8$ supercharge
components is important, in particular the entries with $d=3,4$
and 5. Supersymmetry has already fixed a lot of the structure of
these manifolds, but it is the highest value of $N$ where they are
not yet restricted to symmetric spaces.
\par
The rows in the table are arranged such that the
number of supercharge components is constant. E.g. for $\kappa=16$,
in $d=4$ one has
$4$ charges, each with 4 spinor components, while in $d=10$,
there is one charge with 16 spinor components. These theories can
be related by `dimensional reduction', according to which a
higher-dimensional theory is truncated to a lower-dimensional
one by suppressing the dependence on some of the space-time coordinates.
This relationship forms an important ingredient in the approach
outlined below.
Generally, the procedure of dimensional reduction leaves the
supersymmetries preserved.
\par
{}From now on we  concentrate on the case of 8 supercharge
components, and the relevant supergravity theories have $N=2$
supersymmetry in $d=5$ and $d=4$ and $N=4$ supersymmetry in
$d=3$ space-time
dimensions \cite{GuSiTo,DWVP,dWTNic}. In $d=4$ the
target space factorizes into a \qu\ and a \Ka\ manifold of a
particular type \cite{DWVP}, called {\em special}
\cite{special}. The definition of these special \Ka\ manifolds will
be the subject of section~\ref{ss:defspKa}. The reduction to $d=3$
leads for both factors to a \qu\ manifold. The
interesting results on \qu\ manifolds follow from analyzing how they
emerge in $d=3$ as a result of dimensional reduction from the
first factor in $d=5$ (the `very special real' manifolds) via the
corresponding
$d=4$ special \Ka\ manifold. Actually, we could have started from
$d=6$, but then there are no scalars, and thus no target-space
manifold in the sense described above.
\par
As a
consequence of extended supersymmetry, the target space must be a real, a
K\"ahler or a quaternionic manifold, depending on whether the
supergravity space-time dimension is $d=5$, $d=4$ or $d=3$, respectively.
By means of ordinary dimensional reduction, which preserves
supersymmetry,  one can thus relate real,
complex and quaternionic spaces. More specifically, one defines
two maps, the $\bf r$ and the \cmap, induced by the dimensional
reduction of the corresponding supergravity theories, which act as
\begin{equation}
\Rbar_{n-1} \stackrel{\bf r}{\longrightarrow} \Cbar_n \ ,\qquad
\Cbar_n \stackrel{\bf c}{\longrightarrow} \Hbar_{n+1}\ ,
\end{equation}
Here $n-1$, $n$ and $n+1$ denote the real, complex and
quaternionic dimension of the real, K\"ahler and quaternionic
spaces, respectively. The existence of these maps is based on the
fact that dimensional reduction preserves supersymmetry.
\par
However, the images of these maps do not comprise all the special
\Ka\ and all the \qu\ manifolds. The real manifolds occuring in $d=5$,
$N=2$ supergravity theories coupled to spin-1 fields are
called {\em very special} real manifolds \cite{brokensi}.
Dimensional reduction of the $d=5$ Lagrangians leads to a subclass of
the special \Ka\ manifolds, which will be called very special \Ka\
manifolds. Reduction of the actions containing the special \Ka\
manifolds in $d=4$ to $d=3$ space-time dimensions leads to a
class of quaternionic
manifolds, which will be called `special quaternionic manifolds'.
A subclass of the latter, which constitutes the image of the
\cmap\ acting on the very special \Ka\ manifolds, are the `very
special \qu' manifolds.
\par
All $N=2$ supergravity theories with
spin-1 fields in five space-time dimensions are characterized by
cubic polynomials of $n$ variables.
Hence it is clear that cubic polynomials define a series of
related `very special' real, \Ka\ and quaternionic manifolds. The cubic
polynomials that correspond
to homogeneous manifolds have been classified in \cite{dWVP3}. They
were denoted by $L(q,P)$ and $L(4m,P,\dot P)$, where $q\neq 4m$, $m$,
$P$ and $\dot P$ are
integers restricted by\footnote{At the end of this
text we generalise this to $q=-2$ and $q=-3$ to include
symmetric special \Ka\ and \qu\ manifolds that are not of the very
special type.} $q\geq-1$ and $m,P,\dot P\geq 0$, and they
cover all the homogeneous non-symmetric quaternionic spaces
classified in
\cite{Aleks}.\footnote{Actually, this was already clear from
previous work,
which had revealed that \Al's nonsymmetric spaces were all in the
image of the \cmap\ and that the \cmap\ is in fact closely related to
\Al's classification method.} In
particular, they include the spaces denoted in \cite{Aleks} by
$V(p,q)$ and $W(p,q)$ with $q$ positive, which were related to
$(q+1)$-dimensional Clifford modules. We found, however, that all
the $L(q,P)$ and $L(4m,P,\dot P)$ can be described in a common
framework based on Clifford algebras. For instance, the case of $q=-1$,
where the dimension of the underlying Clifford algebra vanishes,
can be naturally incorporated in the general analysis.
Actually, some of the polynomials did not appear in \Al's
original work. When $q$ is a multiple of 4, $P$ has to be
replaced by the symmetric pair of integers $(P,\dot P)$;
furthermore the case $L(-1,P)$ was missing. Meanwhile these
results have been confirmed in the mathematical literature
\cite{Cortes}.
\par
Here we intend to explain and summarize the results of \cite{ssss}
on the structure of the continuous isometries of these spaces\footnote{
For an alternative summary of these results, see \cite{jrel}.}. These
isometries of the target-space metric extend in fact to
symmetries of the full supergravity action.
\par
The recent new interest of physicists in special geometry and the
symmetry structure of target spaces
is to a large extent motivated by string theory. The
supersymmetric ground states arising from string theory give rise
to an effective field theory of the supergravity type. The moduli
space of these ground states is usually isomorphic to the moduli
space of the restricted target-space manifolds discussed above.
Exploiting these relations and making use of the known
supersymmetry properties of the ground states, one for instance
derives that the moduli space of Calabi-Yau manifolds must exhibit
special geometry \cite{Seiberg,CecFerGir,FerStro,special,DixKapLou,Cand}.
\par
In section~\ref{ss:defspKa} we explain the notion of special \Ka\
manifolds. Then in section~\ref{ss:isospec} we discuss the
isometries of special manifolds. First, in
section~\ref{ss:dualKa}, we specify the isometries
of special \Ka\ manifolds. This will involve the concept of
symplectic transformations, which is a rather important concept in
special \Ka\ manifolds, and at the heart of recent developments. In
section~\ref{ss:isospqu}, the \cmap\ is explained which leads to special
\qu\ manifolds, whose isometries are also discussed. Subsequently we
turn to very special manifolds in section~\ref{ss:isovsp}. We first
explain the \rmap\ and introduce the isometries of very special
real manifolds. This forms the starting point for discussing
the isometries of very special \Ka\ and very special
\qu\ manifolds.
\par
Then we concentrate on homogeneous spaces. We shall see that
\Al's homogeneous non-symmetric spaces are all very special
quaternionic, and in section~\ref{ss:clashomvsp} we give their
classification comprising all known homogeneous \qu\
manifolds. Section~\ref{ss:isovsphom} then discusses their
isometries. It starts from the symmetries of representations of
real Clifford algebras, to determine subsequently the isometries of
the homogeneous very special real manifolds, special \Ka\ manifolds
and \qu\ manifolds. In a final section we summarise all these
results.

\section{Special \Ka\ manifolds} \label{ss:defspKa}
We first briefly introduce the special \Ka\ manifolds in the
context of supergravity. Subsequently we cast the results in a
more abstract form based
on symplectic vectors, which was originally discussed in the context
of the moduli space of Calabi-Yau three-folds.  We close the
section by presenting a
few characteristic examples of special \Ka\ manifolds.
\subsection{Vector multiplets coupled to supergravity }
The scalar sector of the $N=2$ supergravity-Yang-Mills theory
in four space-time dimensions
defines the `special \Ka\ manifolds'. Without
supergravity we have $N=2$ supersymmetric Yang-Mills theory,
whose spinless fields parametrize a similar type of \Ka\
manifolds. The vector potentials, which describe the spin-1
particles, are accompanied by complex scalar fields and doublets
of spinor fields, all taking values
in the Lie algebra associated with the gauge group.
The presence of two independent supersymmetries
implies that the action is encoded
in a holomorphic prepotential $F(X)$, where $X$ denotes the complex
scalar fields \cite{DWVP}.
Two different functions $F(X)$ may correspond to equivalent
equations of motion and to the same geometry. The relation is made
by certain symplectic
transformations that we discuss shortly.
When coupling $n$ of these so-called vector multiplets to
supergravity, one again has a holomorphic prepotential $F(X)$,
this time of $n+1$ complex fields, but now it
must be a {\em homogeneous} function of degree two  \cite{DWVP}.
The physical scalar fields of this system
parameterize an $n$-dimensional complex hypersurface,
defined by the condition that the imaginary part of $X^I\,\bar
F_I(\bar X)$ must be a constant, while the overall phase of the
$X^I$ is irrelevant in view of a local (chiral) invariance.\footnote{%
    Here and henceforth we use the convention where
    $F_{IJ\cdots}$ denote multiple derivatives with respect to $X$ of
    the holomorphic prepotential.
    }
\begin{equation}
i(\bar X^I F_I - \bar F_I X^I) = 1\,.\label{constraint}
\end{equation}
 The embedding of this hypersurface can be described in terms of
$n$ complex coordinates $z^A$ by letting $X^I$ be proportional to
some holomorphic sections $Z^I(z)$ of the projective space
$P\Cbar^{n+1}$ \cite{CdAF}. The $n$-dimensional space parametrized by the
$z^A$ ($A=1,\ldots, n$) is a \Ka\ space; the K\"ahler metric
$g_{A\bar B}=\partial_A\partial_{\bar B}K(z,\bar z)$ follows from the
K\"ahler potential
\begin{eqnarray}
&&K(z,\bar z)=
-\log\Big[i \bar Z^I(\bar z)\,F_I(Z(z)) -i Z^I(z)\,
\bar F_I(\bar Z(\bar z))\Big] \ , \quad \mbox{ where }\label{KP}\\
&& X^I=e^{K/2}Z^I(z)\ ,\qquad\bar X^I=e^{K/2}\bar Z^I(\bar z) \ .
 \nonumber
\end{eqnarray}
The resulting geometry is known as {\em special} \Ka\ geometry
\cite{DWVP,special}. The  curvature tensor associated with this
K\"ahler space satisfies the characteristic relation
\cite{BEC}
\begin{equation}
R^A{}_{\!\!BC}{}^{\!D} = \d^A_{B} \d^D_{C}+\d^A_{C} \d^D_{B}  - e^{2K}
{\cal W}_{BCE}\, \bar {\cal W}{}^{EAD}\, , \label{SKcurvature}
\end{equation}
where
\begin{equation}
{\cal W}_{ABC} =i  F_{IJK}\big(Z(z)\big) \;{\partial
  Z^I(z)\over \partial z^A}  {\partial Z^J(z)\over \partial z^B}
{\partial Z^K(z)\over \partial z^C} \,. \label{defW}
\end{equation}
\par
A convenient choice of inhomogeneous coordinates $z^A$
are the {\em special} coordinates, defined by
\begin{equation}
z^A =X^A/X^0,\qquad A=1,\ldots ,n, \label{defspcoor}
\end{equation}
or, equivalently,
\begin{equation}
Z^0(z)=1\,,\qquad Z^A(z) = z^A\,.
\end{equation}
\par
The kinetic terms of the spin-1 gauge fields in the action are
proportional to the symmetric tensor
\begin{equation}
{\cal N}_{IJ}=\bar
F_{IJ}+2i {{\rm Im}(F_{IK})\,{\rm Im}(F_{JL})\,X^KX^L\over {\rm
Im}(F_{KL})\,X^KX^L} \,.
\label{Ndef}
\end{equation}
This tensor describes the field-dependent generalization of the inverse
coupling constants and so-called $\theta$ parameters.
\par
As mentioned above, the field equations corresponding to two
supersymmetric Yang-Mills actions characterized by different
functions $F(X)$, can be identical up to a symplectic
transformation. In that case the two functions describe
equivalent classical field theories. These symplectic
transformations act as $Sp\left( 2n+2,\Rbar\right)$ rotations on
the vectors $(X^I,F_J)$ and also on the Yang-Mills field-strength
tensors.  However, on
the field strengths they generically rotate electric into magnetic fields
and vice versa. Such rotations, which are called duality
transformations because in four space-time dimensions electric
and magnetic fields are dual to each other (in the sense of
Poincar\'e duality),  cannot be implemented on the vector potentials,
at least not in a local way. Therefore, the
use of these symplectic transformations is only legitimate for zero
gauge coupling constant. From now on we deal
exclusively with Abelian gauge groups. The symplectic
transformations take the form
\begin{equation}
\pmatrix { \tilde X^I \cr \tilde F_I\cr}= {\cal O}
\pmatrix{X^I \cr  F_I\cr} \label{symplXF}
\end{equation}
and generically lead to another prepotential $\tilde F(\tilde X)$,
whose first derivatives with respect to $\tilde X$ correspond to
the $\tilde F_I$. Here ${\cal O}$ is a real $(2n+2)$-by-$(2n+2)$
matrix satisfying the symplectic condition
\begin{equation}
{\cal O}\,\Omega \,{\cal O}^{\rm T} = \Omega\,,
\end{equation}
where
\begin{equation}
\Omega=\pmatrix {0&\unity _{n+1}\cr \noalign{\vskip2mm} -\unity
_{n+1}&0 \cr}\,.
\end{equation}
Hence we may find a variety of descriptions of the same theory
in terms of different functions  $F$. If a symplectic
transformation leads to the same function $F$, then we are
dealing with an invariance of the equations of motion (but not
necessarily of the action as not all transformations can be
implemented locally  on the gauge fields). This invariance
reflects itself in the isometries of the target-space manifold.
We return to this in section~\ref{ss:dualKa}.
\subsection{Symplectic formulation of special geometry}
\label{ss:symplsg}
We now give an alternative and more abstract formulation of
special geometry. This formulation was first given in the context
of a treatment of the moduli space of Calabi-Yau three-folds
\cite{special,FerStro,Cand}. The
connection between these rather different topics hinges on
string theory. The Calabi-Yau manifolds arise as ground-state
configurations for certain string theories, whose low-energy
field theories take the form of an $N=2$ supergravity theory. The
scalar-field sector is locally isomorphic to the moduli space of the
Calabi-Yau manifolds, so that both exhibit special geometry
\cite{Seiberg}. Here we give a
self-contained derivation based on the material presented above.
\par
Based on the previous exposition it makes sense to define a
$(2n+2)$-component vector
$V\equiv(X^I, F_J) \in \Cbar^{2n+2}$, which transforms under
$Sp\left(2n+2,\Rbar\right)$ according to $V\to \tilde V={\cal O} V$.
The constraint (\ref{constraint}) can then be written as\footnote{%
    This inner product arises naturally in the treatment of
    Calabi-Yau manifolds, where $V$ corresponds to the periods of
    a  certain harmonic form. The inner product can be defined in
    the dual cohomology basis and the symplectic group corresponds to
    redefinitions of that basis that leave the inner product (the
    intersection numbers) invariant.    }%
\begin{equation}
\langle \bar V,V\rangle \equiv  \bar V^T \Omega V =-i\ .
\label{constrscale}
\end{equation}
Holomorphic sections $v(z)$, which describe the holomorphic dependence
on the coordinates $z^A$, follow from $V=e^{K/2}v$,
where $K$ is the \Ka\ potential (\ref{KP}), which in this
notation is defined by
\begin{equation}
e^{-K(z,\bar z)}=i\langle \bar v(\bar z), v(z) \rangle \,.
\end{equation}
Here the $(X^I,F_J)$ are the basic
objects; these $2n+2$ quantities are parametrized in terms of the $n$
complex coordinates $z^A$.  We do not impose the condition that
the $F_I$ can be written as the derivatives of a homogeneous
holomorphic function,  so that the dependence of the $F_I$ on
$z^A$ is not necessarily induced via their dependence on the
$X^I$. This starting point is motivated by the fact
that there are situations where the holomorphic prepotential does not
exist, although the sections are well defined \cite{f0art}. The sections $v$
are only defined projectively, i.e., they are uniquely defined modulo
\begin{equation}
v(z) \longrightarrow e^{f(z)}\, v(z)\,. \label{projective}
\end{equation}
Under holomorphic transformations the \Ka\ potential changes
by a \Ka\ transformation
$$
K(z,\bar z)\to K(z,\bar z) - f(z) -\bar f(\bar z)\,,
$$
and the original $\sp$ vector $V$ changes by a phase
transformation
$$
V\to e^{\ft12(f(z)-\bar f(\bar z))}\,V\,.
$$
\par
The holomorphicity of the sections $v$ is expressed by
\begin{eqnarray}
{\cal D}_{\bar A} V &\equiv& \left[ \partial_{\bar A} -\ft12
(\partial_{\bar A} K)\right] V= 0 \,,\nonumber\\
 {\cal D}_{A}\bar  V &\equiv& \left[\partial_A -\ft12
(\partial_A K)\right] \bar V=0 \,.     \label{constr1}
\end{eqnarray}
Here we recognize $\cal D$ as the \Ka\ derivative,
which is covariant with respect to the projective transformations
(\ref{projective}).
For the nonvanishing derivatives of $V$ and $\bar V$ we thus
define
\begin{eqnarray}
{U}_A={\cal D}_A V &\equiv & \left[ \partial_A +\ft12
(\partial_A K)\right] V \,, \nonumber\\
\bar {U}_{\bar A}={\cal D}_{\bar A} \bar V & \equiv &
\left[ \partial_{\bar A} +\ft12
(\partial_{\bar A} K) \right] \bar V\,.
\end{eqnarray}
On the constraint (\ref{constraint}) covariant and ordinary
derivatives coincide, so that application of holomorphic and
anti-holomorphic derivatives leads to
\begin{equation}
\langle \bar V,{U}_A \rangle =\langle \bar{U}_{\bar A},
V \rangle =\langle \bar V,{\cal D}_A{U}_B \rangle=0\,.
\label{VbarU}
\end{equation}
\par
When acting on co- and contravariant vectors the covariant
derivatives contain the Levi-Civita connection associated with the
\Ka\ metric $g_{A\bar B}$, whose nonzero components are
$\Gamma^A_{BC}= g^{A\bar A}\partial_Bg_{C\bar A}$ and their complex
conjugates. Obviously the Levi-Civita connection is not
present in ${\cal D}_{\bar B} U_A$ and we derive
\begin{equation}
{\cal D}_{\bar B} {U}_A= {\cal D}_{\bar B} {\cal D}_A V=
(\partial_{\bar B} \partial_A K) \,V= g_{A\bar B}\,V\ ,
\label{DBbDAV}
\end{equation}
where we used the constraint (\ref{constr1}). The above relation
implies that we are dealing with a so-called \Ka-Hodge manifold.
Furthermore,
using ${\cal D}_{\bar B} {\cal D}_A \langle \bar V,V\rangle =
\partial_{\bar B} \partial_A \langle \bar V,V\rangle  =0$ gives
\begin{equation}
g_{A\bar B} = i \langle {U}_A , \bar {U}_{\bar
B}\rangle \, , \,\mbox{ or, } \quad \langle {U}_A , \bar
{U}^B\rangle = -i\d^B_A\ . \label{metrU}
\end{equation}
As the metric is covariantly constant, one easily establishes
that
\begin{equation}
\langle {\cal D}_AU_B,\bar U_{\bar C}\rangle =0\,,
\end{equation}
making use of (\ref{DBbDAV}) and (\ref{VbarU}).
\par
In (\ref{Ndef}) we defined the tensor ${\cal N}$ in terms of
derivatives of the prepotential $F(X)$. However, it is possible to
find an expression for this tensor without referring to the
prepotential \cite{f0art}. Namely we note the following two
properties, which are direct consequences of the definition
(\ref{Ndef}) and of (\ref{VbarU}),
\begin{equation}
{\cal N}_{IJ} X^J = F_I\,,\qquad {\cal N}_{IJ} {\cal D}_{\bar
A}\bar X^J = {\cal D}_{\bar A}\bar F_I\,,  \label{Ndefnew}
\end{equation}
They enable us to express $\cal N$ as the ratio of two
$(n+1)$-by-$(n+1)$ matrices. In addition, we use that
the matrix $\cal N$ is symmetric. This symmetry requires
one further condition,
\begin{equation}
\langle U_A,U_B\rangle =0 \,,   \label{UU}
\end{equation}
which follows from multiplying the second equation
above by  ${\cal D}_{\bar B}\bar X^I$ and using the symmetry of
$\cal N$.
Acting on (\ref{UU}) with $g^{A\bar C}{\cal D}_{\bar C}$ and
using  (\ref{DBbDAV}) yields
\begin{equation}
\langle V,U_A\rangle=\langle V,{\cal D}_AV\rangle=
\langle V,\partial_AV\rangle = 0 \ . \label{UV}
\end{equation}
When a prepotential $F(X)$ exists, the latter condition is
trivially satisfied\footnote{%
  In discussions on the moduli space of Calabi-Yau three-folds,
  one usually argues that the $F_I$ are locally determined by the
  $X^I$,
  from which the existence of a holomorphic prepotential $F(X)$
  follows \cite{special,FerStro,Cand}. However, we stress that
  examples are known where the
  function $F$ does not exist, simply because the $X^I$ are not
  independent \cite{f0art}.
  Furthermore, in the context of Calabi-Yau manifolds, (\ref{UU})
  follows directly from the decomposition of the cohomology
  basis, while in our treatment it amounts to an
  extra condition (following from the symmetry of $\cal N$).%
  }.
Strictly speaking it follows
only for $n \not= 1$, because \eqn{UU} is trivially satisfied when
$n=1$.
In \cite{whatskg} it is argued that in this case \eqn{UV} should be
imposed as an extra requirement.
Combining (\ref{UU}) and
(\ref{UV}) it follows that
\begin{equation}
\langle V, {\cal D}_AU_B\rangle =\langle V, {\cal D}_A{\cal D}_B
V\rangle =\langle V, \partial_A\partial _B V\rangle = 0 \ .
\label{VDU} \end{equation}

\par
Let us now define the following $(2n+2)$-by$(2n+2)$ matrix $\cal
V$ consisting of the row vectors
\begin{equation}
\cal V = \pmatrix{V\cr \bar U^A \cr \bar V\cr U_A\cr}\ ,
\end{equation}
and transforming from the right under $\sp$. {}From the above
results it is easy to see that $\cal V$ satisfies the symplectic
condition
\begin{equation}
{\cal V}\,\Omega \,{\cal V}^{\rm T} = i\Omega\ ,\label{symplmatrix}
\end{equation}
so that $\cal V$ is isomorphic to an element of $\sp$. Therefore
${\cal V}$ is invertible and we can define $Sp(2n+2)$ connections
${\cal A}_A$ and ${\cal A}_{\bar A}$ such that
\begin{equation}
{\cal D}_A{\cal V}= {\cal A}_A{\cal V}\,, \qquad{\cal D}_{\bar A}
{\cal V}={\cal A}_{\bar A}{\cal V}\,. \label{flatness}
\end{equation}
The values of these two connections can be computed from
${\cal A}_A=i\langle {\cal D}_A{\cal V},{\cal V}^T\rangle\Omega$
and ${\cal A}_{\bar A}=i\langle {\cal D}_{\bar A}{\cal V},{\cal
V}^T\rangle\Omega$. They are completely determined from the
results above, with the exception of the contributions
proportional to the symmetric tensor
\begin{equation}
C_{ABC} \equiv -i\langle{\cal D}_AU_B ,U_C \rangle = -i \langle
V,{\cal D}_A{\cal D}_B{\cal D}_C V\rangle=  -i \langle
V,\partial_A\partial_B\partial_C V\rangle\ , \label{defC}
\end{equation}
the last expression being due to (\ref{UV}) and (\ref{VDU}).
The previous results constrain ${\cal D}_AU_B$ to be only proportional to
$\bar U^A$:
\begin{equation}
{\cal D}_AU_B = C_{ABC}\,\bar U^C\ . \label{DU}
\end{equation}
Incidentally, we note the following equation, which follows from
combining (\ref{DU}) and (\ref{UU}),
\begin{equation}
\langle{\cal D}_A{\cal D}_B V ,{\cal D}_C{\cal D}_D V\rangle=0 \ .
\end{equation}
\par
The connections ${\cal A}$ are now determined and read
\begin{eqnarray}
{\cal A}_A &=& \pmatrix{0&0&0&\d_A^C\cr \noalign{\vskip1mm}
              0&0&\d^B_A&0\cr \noalign{\vskip1mm}
              0&0&0&0\cr \noalign{\vskip1mm}
              0& C_{ABC}&0&0\cr } \,,\nonumber\\[2mm]
{\cal A}_{\bar A} &=& \pmatrix{0&0&0&0\cr \noalign{\vskip1mm}
              0&0&0& \bar C_{\bar A}^{\,BC}\cr \noalign{\vskip1mm}
              0&g_{\bar A C}&0&0\cr \noalign{\vskip1mm}
              g_{\bar AB}&0&0&0\cr } \,.
\end{eqnarray}
Both these connections are nilpotent; the product of more than
three of them vanishes: ${\cal A}_A{\cal A}_B{\cal A}_C{\cal
A}_D=0$ and likewise for the fourth power of the ${\cal A}_{\bar
A}$. This feature was relevant in e.g. \cite{CDFLL,modssym}.
The conditions (\ref{flatness}) imply that the combined
connection consisting of $\cal A$ and the \Ka\ and Levi-Civita
connections must be flat. Calculating the integrability
conditions
\begin{equation}
\left[{\cal D}_C-{\cal A}_C,{\cal D}_{\bar D}-{\cal A}_{\bar D}
\right]{\cal V}=0\ ;\qquad
\left[{\cal D}_C-{\cal A}_C,{\cal D}_D-{\cal A}_D\right]{\cal V}
=0\ ,
\end{equation}
using e.g.
\begin{equation}
\left[{\cal D}_C,{\cal D}_{\bar D}\right] Z_B=Z_A R^A{}_{BC\bar
D}+Z_B (\bar m-m)g_{C\bar D}  \ .
\end{equation}
for a vector $Z_A$ transforming under the \Ka\ symmetry as
$ Z_A \rightarrow  e^{mf(z)+\bar m\bar f(\bar z)}Z_A$,
we obtain the following consequences. First of all the
Riemann curvature (all other curvature components
vanish for a \Ka\ manifold) is given by
\begin{equation}
R^A_{\,BC}{}^{\!D} = \d^A_B\d^D_C+ \d^A_C\d^D_B - C_{BCE}\bar
C^{ADE}\, ,
\end{equation}
which may be compared to (\ref{SKcurvature}).
Secondly, the tensor $C_{ABC}$ satisfies the following two
conditions,
\begin{equation}
{\cal D}_{\bar A}C_{BCD} ={\cal D}_{[A}C_{B]CD} = 0 \ .
\end{equation}
{}From these equations one deduces that ${\cal W}_{ABC}=e^{-K}C_{ABC}$ is
independent of $\bar z$. Furthermore, $C_{ABC}$ can be written
as the third covariant derivative of some scalar function.
\par
This completes the discussion of special geometry. In summary
special \Ka\ manifolds can be defined starting from a complex
$(2n+2)$-component vector $V$ subject to the constraint
\eqn{constrscale}. One demands the existence of symplectic
holomorphic sections $v(z)$ proportional to $V$, such that
$V=\exp({\ft12 K(z, \bar z)})\, v(z)$, and identifies $K$ as
the \Ka\ potential. The holomorphic sections are defined
projectively, as expressed in (\ref{projective}). In addition,
one demands the symmetry of ${\cal N}$ defined by
\eqn{Ndefnew}, or equivalently, imposes \eqn{UU}\footnote{Imposing
instead \eqn{UV} deals also with the $n=1$ case if one wants to
restrict the manifolds to those for which an $N=2$ supergravity
action has been found \cite{whatskg}. }.
Then all the
above result follow.
\par
If, as in the previous subsection, the sections $F_I$ depend on
$X^I$, they should be homogeneous of first degree in $X^I$ and
\eqn{UU} is solved by $F_I=\partial_I F$ for a holomorphic
prepotential $F(X)=\ft12 F_I X^I$.
Then the expression (\ref{defW}) for ${\cal W}_{ABC}$
follows directly from the definition (\ref{defC}) of $C_{ABC}$, and
one can also determine the following expression for $C_{ABC}$
\cite{CDFLL},
\begin{equation}
C_{ABC}= -\ft i2{\cal D}_A{\cal D}_B{\cal D}_C \Big[ \left(
F_{IJ}(X(z,\bar z)) - \bar F_{IJ}(\bar X(z,\bar z))\right)
\,X^I(z,\bar z)\,X^J(z,\bar z) \Big]\,.
\end{equation}
To derive this result one makes use of
$$
{\cal D}_A{\cal D}_B X^I(z,\bar z) = C_{ABC} \,g^{C\bar C} {\cal
D}_{\bar C} \bar X^I(z,\bar z)\ ,
$$
which is implied by (\ref{DU}).

\subsection{Examples of special \Ka\ manifolds}
We give here some examples of functions $F(X)$ and their
corresponding target spaces, which will be useful later on:
\begin{eqnarray}
F=i[(X^0)^2-(X^1)^2]  &\quad& \frac{SU(1,1)}{U(1)}   \\
 F=(X^1)^3/X^0  &\qquad&  \frac{SU(1,1)}{U(1)}   \\
 F=\sqrt{X^0(X^1)^3}   &&  \frac{SU(1,1)}{U(1)}   \\
F=iX^I\eta_{IJ}X^J &&
\frac{SU(1,n)}{SU(n)\otimes U(1)}\\[1mm]
 F=d_{ABC} X^A X^B X^C/X^0 &&  \mbox{`very special \Ka'}
\end{eqnarray}
The first three functions give rise to the manifold $SU(1,
1)/U(1)$. However,
the first one is not equivalent to the other two as the
manifolds have a different value of the curvature \cite{CremVP}.
The latter two are, however, equivalent by means of a symplectic
transformation (\ref{symplXF}). In the fourth example
$\eta$ is a constant non-degenerate real symmetric matrix. In
order that the manifold has a
non-empty positivity domain, the signature of this matrix should be
$(+-\cdots -)$. So not all functions $F(X)$ allow an non-empty
positivity domain. The last example, defined by a real symmetric tensor
$d_{ABC}$, defines a class of special \Ka\ manifolds, which we will
denote as `very special' \Ka\ manifolds. This class of manifolds
is important in the applications discussed below.

\section{Isometries of special manifolds}  \label{ss:isospec}
\subsection{Special \Ka\ : duality transformations} \label{ss:dualKa}
The formulation of special \Ka\ manifolds, as given in
section~\ref{ss:defspKa} shows clearly the possibility of
symplectic reparametrizations, as expressed in \eqn{symplXF}.
Explicitly, such a transformation is of the form
\begin{eqnarray}
\tilde X^I &=& U^I_{\ J}X^J+V^{IJ}F_J\ ,\nonumber\\
\tilde F_I &=& W_{IJ} X^J +Z_I^{\ J} F_J \ .
\label{symplXFX}\end{eqnarray}
When we start from a prepotential $F(X)$, the $F_I$ are the
derivatives of $F$, so that  the first line expresses
the dependence of the new coordinates $\tilde X$ on the old
coordinates $X$. If this transformation is invertible (the full
symplectic matrix itself is always invertible),  the $\tilde F_I$
are again the derivatives of an new function $\tilde F(\tilde X)$
of the new coordinates,
\begin{equation}
\tilde F_I(\tilde X)=\frac{\partial \tilde F(\tilde X)}
{\partial \tilde X^I}\ .
\end{equation}
The integrability condition which implies this statement is equivalent
to the condition that $\pmatrix{U&V\cr W&V}\in \sp$.
Hence we obtain a new, but
equivalent, formulation of the theory, and thus of the
target-space manifold, in terms of the function $\tilde F$.
The manifold was expressed in terms of the vector $(X,F)$, and these
transformations thus give a reparametrization of the same manifold.
The diffeomorphisms are other ways to reparametrize the manifold.
The total group of reparametrizations is thus
\begin{equation}
D_{pseudo}=Diff({\cal M})\times Sp(2(n+1),\Rbar)\ .
\end{equation}
The elements of this group
were called `pseudo-symmetries' in \cite{christoi}.
As explained earlier, the field equations
for the two actions based on functions $F$ and
$\tilde F$ are equivalent, but this equivalence involves duality
transformations, i.e., rotations
between electric and magnetic fields, which cannot be implemented
locally on the underlying vector potentials. Consequently the
relationship cannot be made explicit on the full Lagrangian.
\par
As discussed in section~\ref{ss:symplsg} the $n$ complex
target-space coordinates $z^A$ parametrize the
symplectic sections proportional to $(X^I,F_J)$, which are
subject to the same symplectic transformations \eqn{symplXFX}. In
the simplest case, the sections proportional to the $X^I$ are in
one-to-one correspondence with the coordinates $z^A$, up
to a projective transformation, while the $F_J$ are the derivatives of the
holomorphic prepotential, so that they are determined in terms of
the $X^I$, and therefore in terms of the coordinates $z$. Then the
first line of \eqn{symplXFX} induces a diffeomorphism on the
target-space coordinates. For example, on special coordinates a
symplectic transformation yields
\begin{equation}
z^A= {X^A\over X^0}\quad\longrightarrow \quad\tilde
z^A=\frac{\tilde X^A}{\tilde X^0}\ .  \label{symplspcoor}
\end{equation}
This example clearly exhibits how the symplectic transformations
induce a corresponding diffeomorphism of the coordinates. In a
more general parametrization one encounters a projective term
(here corresponding to the division by $\tilde X^0$) to ensure
that one remains within the initially adopted parametrization.
In the case that there is no function $F(X)$, the $X^I$ are not
independent and one has to choose another subset consisting of
$n+1$ independent components of the symplectic sections.

The question now arises when the symplectic diffeomorphism leaves
the action of the spinless fields invariant, so that it will
constitute an isometry of the target-space manifold. For
simplicity consider the case where a function $F(X)$ exists. By
inverting the previous arguments, it is clear that the
diffeomorphism induces a symplectic transformation on the sections
proportional to $X^I$. However, this does not necessarily mean
that the sections proportional to $F_J$ transform according to
\eqn{symplXFX}. This can only be the case when (up to a quadratic
polynomial with real coefficients)
\begin{equation}
\tilde F(\tilde X)=F(\tilde X)\ . \label{master}
\end{equation}
In other words the symplectic transformations must lead to the
same holomorphic prepotential. Consequently, the corresponding supergravity
theory coincides with the original one, so that in particular
the spin-0 Lagrangian remains the same and the target-space
metric is invariant. When no function $F(X)$ exists, then the
condition \eqn{master} has no meaning. The condition for an
isometry is still that the symplectic transformation on the sections is
correctly induced by the transformations of the coordinates $z$, up to a
projective transformation.

Henceforth the above isometries are called
`duality symmetries', as they are generically accompanied by
duality transformations on the field equations and the Bianchi
identities. However, we are not directly
interested in fields other than the spin-0 ones, but are only
concerned with the symplectic transformations as posssible
isometries of the target-space metric. This changes effectively after
dimensional reduction to $d=3$, as the spin-1 fields (as well as some
components of the space-time metric) are converted to scalar
fields and become part of the target space, as we shall discuss
in the next section.
\par
As infinitesimal transformations, the symplectic transformations are of
the form
\begin{equation} {\cal O}= \unity + {\cal R}\ ;\qquad
{\cal R}=\pmatrix {B&-D\cr C&-B^T\cr}
\qquad \mbox{with}
\begin{array} {l}
C=C^T\ , \\ D=D^T\ .
\end{array} \end{equation}
Then the equation governing duality symmetries is
the infinitesimal form of
(\ref{master}); it reads
\begin{equation}
<V,{\cal R}V>=0 \ . \label{infmaster}
\end{equation}
\par
In the above we started from the symplectic transformations. A
subclass of them are duality symmetries defined by solutions of
\eqn{master} and \eqn{infmaster}. These duality symmetries are
symmetries of the full
supergravity field equations and their action on the scalar
fields defines isometries. The question is whether the
duality symmetries comprise all the isometries of the target
space, i.e. whether
\begin{equation}
Iso({\cal M})\subset Sp(2(n+1),\Rbar)\ . \label{isosubSp}
\end{equation}
We investigated this question in \cite{brokensi} for the very
special \Ka\ manifolds, and found that in this case one does obtain the
complete set of isometries from the symplectic transformations.
For generic special \Ka\ manifolds no isometries have been found
that are not induced by symplectic
transformations, but on the other hand there is no proof that
these do not exist.

\subsection{Special \qu}    \label{ss:isospqu}
We now consider the so-called \cmap\ \cite{CecFerGir} from a
special \Ka\ to a \qu\ manifold. It is induced by reducing
an $N=2$ supergravity action in $d=4$ space-time
dimensions to an action in $d=3$ space-time dimensions, by
suppressing the dependence on one of the (spatial) coordinates.
The resulting $d=3$ supergravity theory can be written in terms
of $d=3$ fields and this rearranges the original fields such that
the number of scalar fields increases from $2n$ to $4(n+1)$, as
is indicated in the table~\ref{tbl:cmap}. The  $4(n+1)$
fields
\begin{table}[ht]\caption{The \cmap\ as dimensional reduction from $d=4$
to $d=3$ supergravity. The number of fields of various spins is
indicated and  names are assigned to the scalar fields in $d=3$.}
\label{tbl:cmap}.
\begin{center}\begin{tabular}{||r|ccc||}\hline
$d=4$ spins& 2 & 1 & 0 \\  \cline{2-4}
numbers &  1  & $n+1$ & 2$n$ \\ \hline
$d=3$ spins & & & \\
2  & 1 &  &  \\
0  & 2 & 2$(n+1)$ & 2$n$ \\
   & $\phi,\sigma$& $A^I,B_I$& $z^A,\bar z^A$\\ \hline
\end{tabular}\end{center}  \end{table}
are denoted by $\phi,\sigma,A^I,B_I,z^A$ and $\bar z^A$, and
parametrize the \qu\ manifold. One may distinguish the following
isometries of this \qu\ manifold \cite{Sabi}:
\begin{itemize}
\item the duality symmetries discussed previously for the
corresponding \Ka\ manifold. $z^A$ are the coordinates of that \Ka\
manifold, and their transformation under these symmetries remains
the same. $(A^I, B_I)$ transform under the duality
transformations as a symplectic
vector, while $\phi$ and $\sigma$ remain inert.
\item shifts and scale transformations which are a consequence of
symmetries of the four-dimensional supergravity theory. We have
the independent shifts of $A^I$, $B_I$ and $\sigma$,
denoted by $\alpha^I$, $\beta_I$ and $\epsilon^+$, and the
scaling denoted by $\epsilon^0$.
\end{itemize}
The latter type of symmetries are called {\em extra} symmetries.
We use this terminology to denote symmetries that can be
understood from the invariances of the
higher-dimensional theory. Such isometries
exist for all manifolds of this type. Below we will also find
so-called {\em hidden symmetries}, for
which there exists no explanation in terms of the underlying
higher-dimensional theory. Such isometries are only present for
particular manifolds, depending on whether certain extra
conditions are satisfied. These conditions have been determined
for all the special \Ka\ and \qu\
manifolds \cite{BEC,dWVP2}.

{}From the algebra of these transformations, we can draw a
root lattice which consists of the root lattice of the duality
symmetry group, extended with one extra dimension as depicted by
the filled circles in table~\ref{tbl:rlsq}.
\begin{table}[ht]\caption{Root lattice of isometries of special \qu\
manifolds. The filled circles represent isometries corresponding
to duality symmetries shifts and scalings, which are present
in all special \qu\ manifolds. The unfilled squares represent
the (hidden) isometries, which only exist for particular manifolds.}
\label{tbl:rlsq}\begin{center}
\setlength{\unitlength}{1mm}
\begin{picture}(150,70)
\put(20,30){\line(1,0){15}}
\put(37,30){\line(1,0){93}}
\put(75,0){\line(0,1){70}}
\put(75,30){\circle*{2}}
\multiput(75,15)(0,10){4}{\circle*{2}}
\multiput(95,8)(0,8){3}{\circle*{2}}
\multiput(95,36)(0,8){3}{\circle*{2}}
\multiput(55,7)(0,8){3}{$\diamond$}
\multiput(55,35)(0,8){3}{$\diamond$}
\put(35,29){$\diamond$}
\put(115,30){\circle*{2}}
\put(70,35){\shortstack[c]{D\\u\\a\\l\\i\\t\\y}}
\put(117,33){\makebox(0,0)[bl]{$\epsilon^+$}}
\put(77.4,26){\makebox(0,0)[bl]{$\epsilon ^0 $}}
\put(97.7,12){\makebox(0,0)[bl]{$\beta_I$}}
\put(97.7,42){\makebox(0,0)[bl]{$\alpha^I$}}
\put(48,11){\makebox(0,0)[bl]{$\hat \beta_I$}}
\put(48,43){\makebox(0,0)[bl]{$\hat \alpha^I$}}
\put(29,33){\makebox(0,0)[bl]{$\epsilon^-$}}
\end{picture}
\end{center}\end{table}
The scale symmetry $\epsilon^0$ is the new element of the Cartan
subalgebra. Together with the $2n+3$ shift symmetries this gives
at least as many additional symmetries as additional coordinates,
as compared
to the original special \Ka\ manifold. These new symmetries provide the
isometries such that a homogeneous \Ka\ manifold gives rise to a
homogeneous \qu\ manifold. Note that the shift and scale symmetries
combined with the solvable part of the duality group constitute  the
solvable algebra of the special \qu\ isometry group.
\par
There can also be
`hidden symmetries' in the special quaternionic manifolds \cite{dWVP2}.
In the root lattice, these appear in the places of the unfilled
squares. Their existence depends on the particular function $F(X)$
we
started from. If all those indicated by $\hat \alpha^I$ and $\hat
\beta_I$ exist, then and only then $\epsilon^-$ exists. This occurs
if and only if the space is symmetric.
\par
As an example consider the ($n=1$) special \Ka\ space
$\frac{SU(1,1)}{U(1)}$. The isometries group $SU(1,1)$ is
represented by $\lambda^0$ as 1-dimensional Cartan subalgebra,
a positive root $\lambda^+$ and a negative one $\lambda^-$.
It leads to the special quaternionic space with isometries given in
table~\ref{tbl:rlG2}.
\begin{table}[ht]\caption{Root lattice of $G_2$.}
\label{tbl:rlG2}
\begin{center}
\setlength{\unitlength}{0.6mm}
\begin{picture}(150,120)
\put(14,60){\line(1,0){9}}
\put(26,60){\line(1,0){110}}
\put(75,5){\line(0,1){110}}
\multiput(75,60)(50,0){2}{\circle*{2}}
\put(75,60){\circle{5}}
\put(100,16.7){\circle*{2}}
\put(100,45.6){\circle*{2}}
\put(100,74.4){\circle*{2}}
\put(100,103.3){\circle*{2}}
\multiput(75,31.1)(0,57.8){2}{\circle*{2}}
\multiput(50,16.7)(0,28.9){4}{\makebox(0,0){$\diamond$}}
\put(25,60){\makebox(0,0){$\diamond$}}
\put(22,52){\makebox(0,0)[br]{$\epsilon^-$}}
\put(78,52){\makebox(0,0)[bl]{$\epsilon ^0$}}
\put(128,52){\makebox(0,0)[bl]{$\epsilon ^+$}}
\put(103,7.7){\makebox(0,0)[bl]{$\beta _0$}}
\put(103,36.6){\makebox(0,0)[bl]{$\beta _1$}}
\put(103,77.4){\makebox(0,0)[bl]{$\alpha^1$}}
\put(103,106.3){\makebox(0,0)[bl]{$\alpha^0$}}
\put(78,22){\makebox(0,0)[bl]{$\lambda^-$}}
\put(78,63){\makebox(0,0)[bl]{$\lambda^0$}}
\put(78,92){\makebox(0,0)[bl]{$\lambda^+$}}
\put(40,7.7){\makebox(0,0)[bl]{$\hat\beta _0$}}
\put(40,36.6){\makebox(0,0)[bl]{$\hat\beta _1$}}
\put(40,77.4){\makebox(0,0)[bl]{$\hat\alpha^1$}}
\put(40,106.3){\makebox(0,0)[bl]{$\hat\alpha^0$}}
\end{picture}
\end{center}     \end{table}
This is an 8 dimensional space: $ \frac{G_2}{SU(2) \otimes SU(2)} $.

\section{Isometries of very special manifolds} \label{ss:isovsp}
The very special manifolds  are defined by a 3-index symmetric tensor
in $n$ dimensions:
 $d_{ABC}$. The real manifold is $n-1$ dimensional, being defined
in terms of $n$ coordinates $h^A$ subject to the constraint
$d_{ABC}h^Ah^Bh^C=1$.
\par
The
reduction of supergravity from $d=5$ to $d=4$induces an `\rmap '
from the very special real manifolds to very special \Ka\
manifolds. This is depicted in table~\ref{tbl:rmap}.
\begin{table}[ht]\caption{The \rmap\ induced by dimensional
reduction from $d=5$ to
$d=4$ supergravity. The number of fields of integer spins is indicated.}
\label{tbl:rmap}
\begin{center}\begin{tabular}{||r|ccc||}\hline
$d=5$ spins& 2 & 1 & 0 \\  \cline{2-4}
numbers &  1  & $n$ & $n-1$ \\ \hline
$d=4$ spins & & & \\
2  & 1 &  &  \\
1  & 1 & $n$ &  \\
0  & 1 & $n$ & $n-1$ \\ \hline
\end{tabular}\end{center}  \end{table}
This map leads to a scalar manifold with $2n$ real coordinates or
$n$ complex ones. It can be followed by a \cmap\ to
obtain $n+1$ quaternions, as was explained before.

\subsection{Isometries of very special real manifolds}
Let us now consider the isometries of these manifolds \cite{GuSiTo}.
They are determined by the matrices $\tilde B$ which satisfy\footnote{
$(\cdots )$ indicates a symmetrisation in the corresponding
indices with `weight one', e.g. $V_{(A}W_{B)}=\ft12 \left(
V_AW_B+V_BW_A\right) $.}
\begin{equation}
d_{D(AB}\tilde B^D{}_{C)} = 0\ . \label{isosrm}
\end{equation}
\subsection{Isometries of very special \Ka\ manifolds} \label{ss:isovsk}
The duality transformations of the corresponding
\Ka\ manifold are defined by the symplectic matrix
\begin{eqnarray}
\pmatrix {B& -D\cr C & -B^T\cr} \ ; \qquad &&B=\pmatrix{\beta & a_B \cr
 b^A &\tilde B^A_{\;B} +\frac{1}{3}\beta \,\delta^A_{\,B}\cr}  \nonumber \\
                                   && C= \pmatrix{0 & 0 \cr
 0 & 6 d_{ABC}\,b^C\cr} \nonumber \\
                                   && D=  \pmatrix{0 & 0 \cr
 0 & 6 e^{2K} d^{ABC}\,a_C} \ ,
\end{eqnarray}
where $d^{ABC}$ is $d_{ABC}$ where the indices are raised with the
metric.
They consist of the following isometries:
\begin{itemize}
\item $\tilde B$ are the solutions of \eqn{isosrm}.
\item $\beta\, , b^A$ constitute $n+1$ extra isometries
originating from the symmetries of the $d=5$ theory. Hence these
isometries exist for all very special \Ka\ manifolds.
\item Hidden symmetries
exist for those independent parameters $a_A$ for which
$R^A{}_{\!BC}{}^D\, a_D$ is constant (in special coordinates).
\end{itemize}
In the root space the isometries appear generically as indicated in
table~\ref{tbl:isovsk}. Compared to the very special real space,
the corresponding \Ka\ space has at least as many additional
isometries as it has additional coordinates
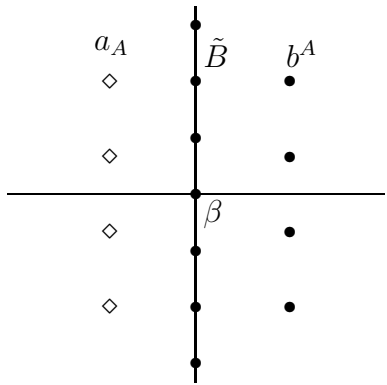
\begin{table}[ht]\caption{Root lattice of the isometries in very special
\Ka\ manifolds.}
\label{tbl:isovsk}
\begin{center}
\setlength{\unitlength}{0.5mm}
\begin{picture}(100,100)
\put(0,50){\line(1,0){100}} \thicklines
\put(50,0){\line(0,1){100}}
\multiput(25,18)(0,20){4}{$\diamond$}
\multiput(75,20)(0,20){4}{\circle*{3}}
\multiput(50,5)(0,15){7}{\circle*{3}}
\put(23,87) {\makebox(0,0)[bl]{$a_A$}}
\put(74,84) {\makebox(0,0)[bl]{$b^A$}}
\put(52,84) {\makebox(0,0)[bl]{$\tilde B$}}
\put(52,41) {\makebox(0,0)[bl]{$\beta$}}
\end{picture}
\end{center} \end{table}

\subsection{Isometries of very special  \qu\ manifolds}
In the very special \qu\ manifold (the image under the \cmap\ of a
very special \Ka\ manifold), one finds the following isometries:
\begin{itemize}
\item all the isometries of the very special \Ka\ manifold.
\item $\alpha^I\, ,\beta_I\, ,\epsilon^0\, ,\epsilon^+$ constitute
$2n+4$ extra isometries, as followed already from the general
statement made for special \qu\ manifolds.
\item $\hat \beta_0$ is an isometry present for all
very special  \qu\ manifolds.
\item Additional hidden symmetries can exist. There are symmetries
$\hat \beta_A$ under the same condition as for the existence of the $a_A$
in section~\ref{ss:isovsk}. If some of the $\hat\b_A$ are
realized, then there are additional ones characterized by
$\hat \alpha^A=d^{ABC}\hat \beta_B\hat \beta_C$ (for the exact
conditions see section 4.3 of \cite{ssss}). If the curvature is
covariantly constant
in all directions (a symmetric manifold) then all $\hat \beta_A$ and
$\hat \alpha^A$  exist and so do the isometries $\hat \alpha^0$
and $\epsilon^-$.
\end{itemize}
The above  isometries give rise to the root lattice depicted in
table~\ref{tbl:rlvsqu}.
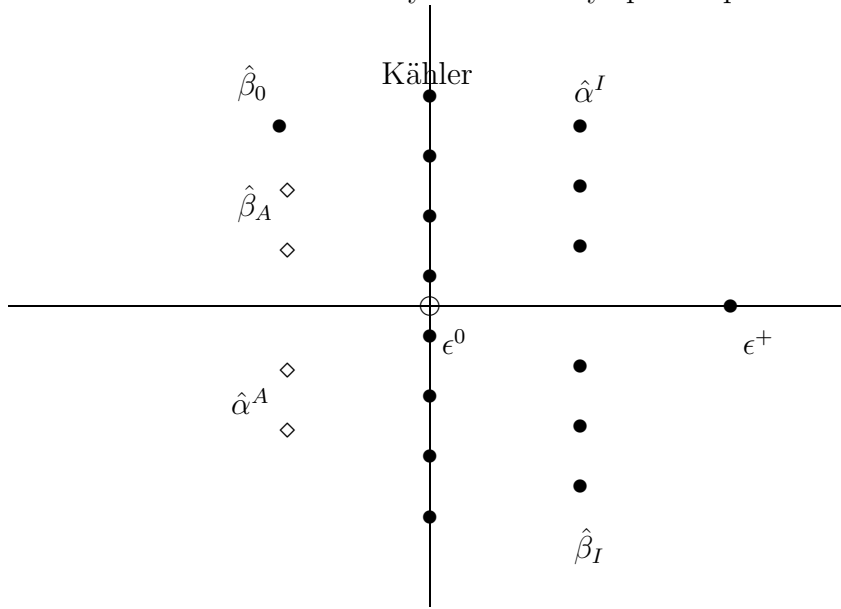
\begin{table}[ht]\caption{Root lattice of isometries of non-symmetric
very special \qu\ manifolds.}
\label{tbl:rlvsqu}
\begin{center}
\setlength{\unitlength}{0.8mm}
\begin{picture}(140,100)(-20,0)
\put(-20,50){\line(1,0){140}}
\put(50,0){\line(0,1){100}}
\multiput(25,28)(0,10){2}{$\diamond$}
\multiput(25,58)(0,10){2}{$\diamond$}
\put(25,80){\circle*{2}}
\multiput(75,20)(0,10){3}{\circle*{2}}
\multiput(75,60)(0,10){3}{\circle*{2}}
\multiput(50,15)(0,10){8}{\circle*{2}}
\put(50,50){\circle{3}}
\put(100,50){\circle*{2}}
\put(18,84) {\makebox(0,0)[bl]{$\hat \beta_0$}}
\put(18,64) {\makebox(0,0)[bl]{$\hat \beta_A$}}
\put(17,32) {\makebox(0,0)[bl]{$\hat \alpha^A$}}
\put(74,7) {\makebox(0,0)[bl]{$\hat \beta_I$}}
\put(74,84) {\makebox(0,0)[bl]{$\hat \alpha^I$}}
\put(42,87) {\makebox(0,0)[bl]{\Ka }}
\put(52,42) {\makebox(0,0)[bl]{$\epsilon^0$}}
\put(102,42) {\makebox(0,0)[bl]{$\epsilon^+$}}
\end{picture}
\end{center}  \end{table}

\section{Classification of homogeneous very special manifolds}
\label{ss:clashomvsp}

In his classification \Al\ obtained  non-symmetric homogeneous
quaternionic manifolds \cite{Aleks} .
It was shown by Cecotti \cite{Cecotti} that all of these are in
fact in the category of very special \qu\ manifolds. This led us to
investigate all the homogeneous very special \qu\ manifolds
\cite{dWVP3}. We proved a theorem \cite{ssss} that there is a
one-to-one mapping between all the homogeneous special \qu\ manifolds and
the homogenous special \Ka\ manifolds\footnote{The theorem
presupposes that the homogeneity of the special \Ka\ manifold is due
to isometries satisfying \eqn{isosubSp}. For very special \Ka\
manifolds it was shown that all iosometries are contained in the
duality transformations \cite{brokensi}. For other special
manifolds no counter examples are known.}, and similarly that
there is a one-to-one mapping between all the homogeneous very special
\Ka\ and the homogeneous very special real manifolds. Therefore
one can start from the very special real manifolds,
characterised by the symmetric tensor $d_{ABC}$.
\par
First we adopt a so-called canonical parametrization \cite{GuSiTo},
in which the tensor takes the form: (with $A=1$ or $a=2,\ldots n$)
\begin{equation}
d_{111}=1\ ;\qquad d_{11a}=0\ ;\qquad d_{1ab}=-\ft12 \delta _{ab}\ .
\end{equation}
Imposing that solutions of \eqn{isosrm} provide transitive isometries,
leads to the equation
\begin{equation}
d_{e(ab}\,d_{cd)e}-\ft12 \,
\delta_{(ab}\,\delta_{cd)}=d_{e(ab}\,A_{c)e;d}
\end{equation}
where $A$ is an arbitrary tensor, antisymmetric in
its first two indices. This condition can be solved after an
elaborate series of steps and we arrived at the following general
solution
(after some redefinitions, so that we are no longer in the canonical
parametrization) \cite{dWVP3}. First, decompose the indices $A$
into $A= 1, 2, \mu, i$, with $\mu=1,\ldots, q+1$ and $i=1,\ldots, r$,
so that $n= 3+q+r$. Hence we assume $n\geq 2$. The result can
then be expressed as follows,
\begin{equation}
d_{ABC}\,h^Ah^Bh^C = 3\Big\{ h^1\,\big(h^2\big)^2 -h^1\,
\big(h^\mu\big)^2 -h^2\,\big(h^i\big)^2
+\gamma_{\mu ij}\,h^\mu\, h^i\,h^j \Big\}\ .\label{soldhom}
\end{equation}
Here the coefficients $\gamma_{\mu ij}$ are $q+1$ real $r\times r$
gamma matrices that generate a real Clifford algebra of positive
signature (${\cal C}(q+1,0)$).
Therefore the solutions are completely classified by specifying a
representation of this Clifford algebra. The irreducible representations
of these Clifford algebras are unique, except for $q=4m$, when there
are two inequivalent ones, which are, however, interchangeable.
Therefore we can denote the solution by
$L(q,P)$ for $q\neq 4m$ and $L(4m,P,\dot P)$, where
$P$ is the number of irreducible representations, and
\begin{equation} L(4m,P,\dot P)=L(4m,\dot P,P)\ .\end{equation}
\par
When
\begin{equation}
 \left( \gamma _\mu
\right) _{(ij} \left( \gamma _\mu \right) _{k\ell )} =
\delta_{(ij}\delta _{k\ell )}\ ,
\end{equation}
the maximal number of hidden isometries is realized for the real,
\Ka\ and \qu\ case. The corresponding spaces are then symmetric
and are listed in table~\ref{tbl:symvs}.
\begin{table}[ht]\caption{Symmetric very special manifolds}
\label{tbl:symvs}
\begin{center}
\begin{tabular}{|l|ccc|}
\hline
&real & K\"ahler & quaternionic \\
\hline
$L(-1,0)$&$SO(1,1)$&$\left[\frac{SU(1,1)}{U(1)}\right]^2$&$\frac{SO(3,4)}{(
S U ( 2 ) ) ^ 3 } $ \\
$L(0,P)$&$ SO(1,1)\otimes \frac{SO(P+1,1)}{SO(P+1)}$&
$\frac{SU(1,1)}{U(1)}\otimes\frac{SO(P+2,2)}{SO(P+2)\otimes SO(2)}$ &
$\frac{SO(P+4,4)}{SO(P+4)\otimes SO(4)} $\\
$L(1,1)$&
$\frac{S\ell(3,R)}{SO(3)}$&$\frac{Sp(6)}{U(3)
}$&$\frac{F_4}{USp(6)\otimes SU(2)}$\\
$L(2,1)$&
$\frac{S\ell(3,C)}{SU(3)}$&$\frac{SU(3,3)}{SU(3)\otimes
SU(3)\otimes U(1)}$&$\frac{E_6}{SU(6)\otimes SU(2)}$\\
$L(4,1)$&
$\frac{SU^*(6)}{Sp(3)}$&$\frac{SO^*(12)}{SU(6)\otimes
U(1)}$&$\frac{E_7}{\overline{SO(12)}\otimes SU(2)}$\\
$L(8,1)$&
$\frac{E_6}{F_4}$&$\frac{E_7}{E_6\otimes
 U(1)}$&$\frac{E_8}{E_7\otimes SU(2)}$\\
\hline \end{tabular}
\end{center}   \end{table}
However, this table does not comprise all the symmetric very
special spaces. Also the real very special real manifolds $L(-1,P)$ are
symmetric, as we shall discuss at the end of this section. One
may now consider the question
whether table~\ref{tbl:symvs} contains all symmetric special \Ka\ and
\qu\ spaces.
For the special \Ka\ manifolds a complete classification of the
symmetric manifolds was given in \cite{CremVP}. As it turns out,
apart from those already found, there is the image under the
\rmap\ of an empty very
special real manifold (it corresponds to the pure supergravity action
in $d=5$, i.e. without scalar fields), and there are the complex projective
spaces. The latter are not of the very special type.
Similarly the symmetric quaternionic spaces that were not yet
mentioned, are those in the image under the \cmap\ of the
special \Ka\ manifolds just mentioned and of the empty special \Ka\
manifold (pure supergravity in $d=4$); in addition there are the
quaternionic
projective spaces. This leads to table~\ref{tbl:homsp}.
\begin{table}[ht]\caption{Homogeneous manifolds.
In this table, $q$, $P$, $\dot P$ and $m$ denote positive integers or
zero, and $q\neq 4m$. SG denotes
an empty space, which corresponds to supergravity models without
scalars.
Furthermore, $L(4m,P,\dot P)=L(4m,\dot P,P)$. The horizontal
lines separate spaces of different rank. The first non-empty space in
each column has rank 1. Going to the right or down a line increases the
rank by~1. The manifolds indicated by a $\star$ were not known before
our classification, except for the cases $L(0,P,\dot P)$.}
\label{tbl:homsp}
\begin{center}
\begin{tabular}{|l|ccc|}
\hline
&real & K\"ahler & quaternionic \\
\hline &&&\\[-3mm]
$L(-3,P)$&&   & $\frac{USp(2P+2,2)}{USp(2P+2)\otimes SU(2)} $
     \\[2mm]
$SG_4$&&  SG       &$\frac{U(1,2)}{U(1)\otimes U(2)} $
    \\[2mm]
\hline&&&\\[-3mm]
$L(-2,P)$&&$\frac{U(P+1,1)}{U(P+1)\otimes U(1)}$
    &$\frac{SU(P+2,2)}{SU(P+2)\otimes SU(2)\otimes U(1)} $
        \\[2mm]
$SG_5$&SG  & $\frac{SU(1,1)}{U(1)}$
    &$\frac{G_2}{SU(2)\otimes SU(2)}$  \\[2mm]
\hline&&&\\[-3mm]
$L(-1,P)$&$\frac{SO(P+1,1)}{SO(P+1)}$& $\star$ & $\star$ \\[2mm]
\hline&&&\\[-3mm]
$L(4m,P,\dot P)$& $\star$ & $\star$& $\star$ \\[2mm]
$L(q,P)$&$X(P,q)$&$H(P,q)$&$V(P,q)$\\[2mm]
\hline \end{tabular}
\end{center} \end{table}
Note that we extended the notation $L(q,P)$ to include $q=-2$ and $q=-3$,
which have no meaning as Clifford algebra ${\cal C}(q+1,0)$.
In these cases also the real manifold does not exist, which brings
us outside the framework on which this notation was based.
We further elucidate this in the next section.
\par
Note that we started from the equation \eqn{isosrm}, which
is the condition for symmetries of the full
$d=5$ supergravity theory, leading
to what were called duality symmetries in the \Ka\ case. We mentioned
already that for very special \Ka\ manifolds all the isometries are
obtained as duality transformations, i.e. \eqn{isosubSp} holds. On
the other hand, the case $L(-1,P)$ exhibits target-space
isometries that are not an invariance of the full $d=5$ supergravity
theory. These additional isometries promote the target space to a
symmetric space.  When applying the \rmap, the non-invariant
sector of the supergravity theory
become relevant for the scalar sector of the very special \Ka\
manifold, and therefore these transformations do not constitute
isometries of the very special \Ka\ manifold. As a result we have
that the real manifold $L(-1,P)$ is a symmetric space ($
\frac{SO(P+1,1)}{SO(P+1)}$), while the corresponding
\Ka\ and \qu\ spaces are non-symmetric (but still homogeneous
because the solvable subalgebra corresponds to a transitive group
of motions, which leaves the full supergravity action invariant).

\section{Isometries of very special homogeneous manifolds}
\label{ss:isovsphom}

We now apply the above results for very special manifolds
to the homogeneous manifolds.

\subsection{Symmetries of representations of real Clifford algebras}

We start from the symmetries of the Clifford algebra. They
consist of the rotation group $SO(q+1)$ (or rather its cover
group) and the matrices $S$ satisfying
\begin{equation}
[\gamma _\mu ,S]=0\ ;\qquad S=-S^T\ .
\end{equation}
This defines ${\cal S}_q(P, \dot P)$, the metric-preserving group in
the centralizer of the Clifford algebra. On an irreducible
representation, Schur's lemma implies that the solution of the
first condition is just the division algebra $\Rbar , \ \Cbar$ or
$\Hbar $. On reducible representations we obtain the general matrices
$\Rbar (P)$, $\Cbar(P)$ or $\Hbar (P)$. The metric-preserving
property (second condition) reduces this to $SO(P), \ SU(P)$ or
$USp(2P)$. The real Clifford algebras, their dimensions and the group
${\cal S}_q$ are given in table~\ref{Cliffp0} \cite{CliffordR}.
\begin{table}[htb]
\begin{center}
\begin{tabular}{||c|c|c|l||}\hline
$q$  &${\cal C}(q+1,0)$& ${\cal D}_{q+1}$&${\cal S}_q(P,\dot P)$
\\ \hline
$-1$ &$\Rbar$    &1         &$SO(P)$     \\
0    &$\Rbar\oplus \Rbar $&1&$SO(P)\otimes SO(\dot P)$ \\
1    &$\Rbar(2)$ &2         &$SO(P)$     \\
2    &$\Cbar(2)$ &4         &$U(P)$     \\
3    &$\Hbar(2)$ &8         &$USp(2P)$\\
4    &$\Hbar (2)\oplus \Hbar (2)$&8&$USp(2P)\otimes USp(2\dot P)$\\
5    &$\Hbar(4)$ &16&$USp(2P)$   \\
6    &$\Cbar(8)$ &16&$U(P)$     \\
7    &$\Rbar(16)$&16&$SO(P)$     \\
$n+8$ &  $\Rbar(16)\otimes{\cal C}(n+1,0)$&16 ${\cal D}_n$ &
as for $q=n$\\
\hline
\end{tabular}
\end{center}
\caption{Real Clifford algebras ${\cal C}(q\!+\!1,0)$. Here
${\bf F}(n)$ stands for $n\times n$
matrices with entries over the field
$\bf F$, while ${\cal D}_{q+1}$ denotes the
real dimension of an irreducible representation of
the Clifford algebra. ${\cal S}_q(P,\dot P)$ is the metric
preserving group in the centralizer of the Clifford algebra in
the $(P+\dot P) {\cal D}_{q+1}$-dimensional representation.}
\label{Cliffp0}
\end{table}

\subsection{Isometries of homogeneous very special real manifolds}
Here we have the solutions of \eqn{isosrm} for the $d$-symbols
given by \eqn{soldhom}. First of all, a scaling symmetry $\lambda $
is obvious from the form of $d_{ABC}$.
Other solutions are in eigenspaces of of this symmetry
\begin{equation}
{\cal X} =   {\cal X}_0\oplus {\cal X}_{3/2} \end{equation}
\begin{eqnarray}
 {\cal X}_0 &=& \lambda  \oplus so(q+1,1) \oplus {\cal S}_q(P,\dot
 P)\nonumber\\
{\cal X}_{3/2}&= &(spinor,vector)\ ,
\end{eqnarray}
where $spinor$ denotes a spinor representation of
${\cal C}^+(q+1,1)\simeq {\cal C}(q+1,0)$
(of dimension ${\cal D}_{q+1}$).
In case the space is symmetric, also ${\cal X}_{-3/2}$ appears with
the same assignment under ${\cal X}_0$ as ${\cal X}_{3/2}$.
\par
The isotropy group is always
\begin{equation}
H= SO(q+1) \otimes {\cal S}_q(P,\dot P)\ .
\end{equation}
Guided by the $SO(q+1,1)$ in the isometry group, we can rewrite the
result \eqn{soldhom} in an $SO(q+1,1)$ invariant form:
\begin{equation}d_{ABC}h^A\,h^B\,h^C=-\eta_{MN}\,h^Mh^Nh^1 + \gamma_{M\,ij}
\, h^Mh^ih^j  \end{equation}
where $M$ takes $q+2$ values (the value 2 or $\mu$ from before),
$\eta_{MN}$ is the $SO(q+1,1)$ metric, and
\begin{equation}
\gamma_M= \pmatrix{0 & \gamma_{M\,ik}\cr
       \noalign{\vskip 1mm}
     \tilde   \gamma_M{}^{jl} & 0 \cr }
\end{equation}
are the corresponding $\gamma$-matrices.

\subsection{Isometries of homogeneous special \Ka\ manifolds}
As isometries of the homogeneous very special \Ka\ manifold we find:
\begin{itemize}
\item the isometries discussed for the real manifold.
\item the $n+1$ extra isometries which appear for all very special \Ka\
manifolds, of which $\beta $ appears in the Cartan subalgebra.
\item the condition for hidden isometries has
always $q+2$ solutions: $a_M$.
\end{itemize}
This leads to the root diagram of table~\ref{tbl:isohomvsk}.
\begin{table}[ht]\caption{Isometries of non-symmetric
homogeneous special \Ka\ manifolds.}
\label{tbl:isohomvsk}
\begin{center}
\setlength{\unitlength}{0.6mm}
\begin{picture}(200,70)
\put(20,30){\line(1,0){110}}
\put(75,0){\line(0,1){70}}
\put(75,30){\circle*{2}}
\put(75,30){\circle{5}}
\put(75,60.6){\circle*{2}}
\multiput(103.9,9.6)(0,30.6){3}{\circle*{2}}
\put(65,33){\makebox(0,0)[bl]{${\cal X}_0$}}
\put(65,63.6){\makebox(0,0)[bl]{${\cal X}_{\frac{3}{2}}$}}
\put(77.4,21){\makebox(0,0)[bl]{$\beta $}}
\put(106.3,6.6){\makebox(0,0)[bl]{$b^M$}}
\put(106.3,37.2){\makebox(0,0)[bl]{$b^i$}}
\put(106.3,67.8){\makebox(0,0)[bl]{$b^1$}}
\put(46.1,50.4){\makebox(0,0){\circle*{2}} }
\put(43.7,47.4){\makebox(0,0)[br]{$a_M$}}
\end{picture}
\end{center}  \end{table}
By rotating the axes of this figure, the root diagram contains again
only non-negative roots. As root space we obtain
\begin{eqnarray}
&&{\cal W}={\cal W}_0 \oplus  {\cal W}_1 \oplus  {\cal W}_2\ ,
\nonumber\\
&&{\cal W}_0= \lambda '\oplus so(q+2,2)\oplus {\cal
S}_q(P,\dot P) \ ,\nonumber\\
&&{\cal W}_1= (1, spinor, vector)\
,\nonumber\\
&&{\cal W}_2= (2,0,0)\ ,
\end{eqnarray}
In ${\cal W}_1$ appears a spinor representation of
\begin{equation}{\cal C}^+(q+2,2)\simeq{\cal
C}(q+2,1)\simeq {\cal C}(q+1,0)\otimes \Rbar (2)\end{equation}
This representation comprises the roots in ${\cal X}_{\frac{3}{2}}$
and the $b^i$. It thus contains $P$ or $P+\dot P$ spinors of dimension
$2{\cal D}_{q+1}$
each. The isotropy group is
\begin{equation}
H= SO(q+2)\otimes U(1) \otimes {\cal S}_q(P,\dot P)\ .
\end{equation}
Again in the case that the manifold is symmetric, we find also ${\cal
W}_{-1}$ and ${\cal W}_{-2}$ with the same assignments as ${\cal W}_1$
and ${\cal W}_2$.
\par
Note that in this classification of the isometries, $q$ appears
always in the linear combination $q+2$. Therefore one may
consider the case $q=-2$,
which made no sense for the real spaces. Then it turns out that we
obtain the isometries of the special \Ka\ manifolds
$\frac{U(P+1,1)}{U(P+1)\otimes U(1)}$, which are, however, not
very special. This motivates the assignment $L(-2,P)$, used in
table~\ref{tbl:homsp}.

\subsection{Isometries of homogeneous quaternionic manifolds}

For the  homogeneous special quaternionic the isometries are
\begin{itemize}
\item isometries of the special \Ka\ manifold.
\item $2n+5$ extra isometries as for all very special quaternionic
manifolds, of which $\epsilon ^0 $ is the extra element in the Cartan
subalgebra.
\item $q+2$ isometries $\hat \beta_M$ and one isometry $\hat \alpha ^1$.
\end{itemize}
These isometries appear in a root diagram as indicated in
table~\ref{tbl:isohomq}.
\begin{table}[ht]\caption{Isometries of non-symmetric homogeneous
very special quaternionic manifolds.}
\label{tbl:isohomq}
\begin{center}
\setlength{\unitlength}{1mm}
\begin{picture}(130,70)
\put(20,30){\line(1,0){110}}
\put(75,0){\line(0,1){70}}
\put(75,30){\circle*{2}}
\put(75,30){\circle{5}}
\put(75,45){\circle*{2}}
\put(75,60){\circle*{2}}
\multiput(95,15)(0,15){3}{\circle*{2}}
\put(115,30){\circle*{2}}
\put(55,45){\circle*{2}}
\put(65,33){\makebox(0,0)[bl]{${\cal W}_0$}}
\put(65,48){\makebox(0,0)[bl]{${\cal W}_1$}}
\put(70,63){\makebox(0,0)[bl]{$1$}}
\put(77.4,26){\makebox(0,0)[bl]{$\epsilon _0 $}}
\put(97.7,12){\makebox(0,0)[bl]{$v$}}
\put(97.7,26){\makebox(0,0)[bl]{$s$}}
\put(117.7,26){\makebox(0,0)[bl]{$1$}}
\put(97.7,42){\makebox(0,0)[bl]{$v$}}
\put(49,42){\makebox(0,0)[bl]{$v$}}
\end{picture}
\end{center}   \end{table}
Again by a rotation of the axes all roots become non-negative
and can be combined as follows:
\begin{eqnarray}
&&{\cal V}={\cal V}_0 \oplus  {\cal V}_1 \oplus  {\cal V}_2\ ,
\nonumber\\
&&{\cal V}_0= \epsilon '\oplus so(q+3,3)\oplus {\cal
S}_q(P,\dot P) \ ,\nonumber\\
&&{\cal V}_1= (1, spinor, vector)\
,\nonumber\\
&&{\cal V}_2= (2,vector,0)\ .
\end{eqnarray}
The isotropy group is
\begin{equation}
H= SO(q+3)\otimes SU(2) \otimes {\cal S}_q(P,\dot P)\ .
\end{equation}
The dimension of the irreducible spinor representations is now
$4{\cal D}_{q+1}$.
For the symmetric manifolds we have also ${\cal V}_{-1}$ and
${\cal V}_{-2}$ with similar assignments as ${\cal V}_{1}$ and
${\cal V}_{2}$.
\par
For the isometries of the \qu\ manifolds the linear combination
$q+3$ appears everywhere, so we can consider also the case $q=-3$.
Then we obtain the root lattice of the quaternionic homogeneous
(and symmetric) spaces, which were called `of type I' in \cite{Aleks}.
These are the quaternionic projective spaces
$\frac{USp(2P+2,2)}{USp(2P+2)\otimes USp(2)}$, which thus deserve the
name $L(-3,P)$ as indicated in table~\ref{tbl:homsp}.

\section{Summary}
The notion of special \Ka\ manifolds induces also a notion of `{\it
special quaternionic manifolds}', which are those manifolds appearing
in the image of the \cmap. Similarly, from the very special
real manifolds (characterised by a symmetric tensor
$d_{ABC}$), very special \Ka\ manifolds are induced as the image of the
\rmap\ and very special quaternionic manifolds as the image of
the \crmap.
\par
The {\it homogeneous} very special real manifolds, special \Ka\ manifolds
and quaternionic manifolds are classified as $L(q,P)$ for $q\neq 4 m$
or $L(4m,P, \dot P)=L(4m, \dot P,P)$, where $P$ and $\dot P$ are
non-negative integers. $q$ is also an integer with $q\geq -1$ for
the very special real manifolds, $q\geq -2$ for the special \Ka\
manifolds and $q\geq -3$ for the \qu\ manifolds. These
classifications are related to spinor representations of resp.
$SO(q+1,1)$, $SO(q+2,2)$ and $SO(q+3,3)$. All those are related by
equivalences
\begin{equation}
{\cal C}(p,q)\otimes \Rbar(2)\simeq {\cal C}(p+1,q+1)  \ ;\qquad
{\cal C}^+(q+r,r)\simeq {\cal C}(q+r,r-1)\ .
\end{equation}
The $d$-symbols of the very special manifolds were defined in terms
of realizations of  real positive-definite
Clifford algebras ${\cal C}(q+1,0)$.
The special \Ka\ manifold $SU(1,1)/U(1)$, the image under the \rmap\ of
the empty very special real manifold, its \qu\ image under the
\cmap,
$\frac{G_2}{SU(2)\otimes SU(2)}$, and the \qu\ manifold
$\frac{U(1,2)}{U(1)\otimes U(2)}$, image under the \cmap\ of the empty
special \Ka\ manifold, are not contained in this scheme. On the
other hand, this set-up led to several homogeneous \qu\ manifolds
that did not occur in \Al's classification \cite{Aleks}.
\par
For both the real, the \Ka\ and the \qu\ case, the isometries
exhibit a grading with respect to
one generator, denoted by $\lambda $. For the non-symmetric
homogeneous manifolds the root space is
\begin{eqnarray}
&& G=G_0 \oplus  G_1 \oplus  G_2\nonumber \\
&& G_0=\lambda\otimes  {\cal O}\otimes
{\cal S}_q(P,\dot P)\nonumber \\ && G_1=(1,spinor,vector)\ ,
\end{eqnarray}
where ${\cal S}_q(P)= SO(P)$ or $U(P)$ or $USp(2P)$
(see table~\ref{Cliffp0}), $spinor$ denotes a spinor representation
of the group ${\cal O}$, which is of dimension ${\cal D}_{q+1}$ for
real manifolds, $2{\cal D}_{q+1}$ for the \Ka\ case, or
$4{\cal D}_{q+1}$ for the \qu\ case. The group ${\cal O}$ and
eigenspace $G_2$ are given by
\begin{equation}
\begin{array}{||l||c|c|c||} \hline
& \mbox{real} & \mbox{\Ka } & \mbox{quatern.} \\  \hline
{\cal O}& SO(q+1,1) & SO(q+2,2) & SO(q+3,3)\\
G_2 & - & singlet & (vector,1) \\ \hline
\end{array}
\end{equation}
In symmetric spaces also $G_{-1}$ and $G_{-2}$ occur.
\vspace{ 1cm}
\par
\noindent{ \bf Acknowledgements}\vspace{0.3cm}
\par
We thank D. \Al\ and V. Cort\'es for interesting discussions. We
have profited from a stimulating meeting in Trieste, for which
we thank the organizers.
\par
This work was carried out in the framework of the
project "Gauge theories, applied supersymmetry and quantum
gravity", contract SC1-CT92-0789 of the European Economic
Community.


\begin{thebibliography}{99}
\bibitem{Aleks} D.V. \Al , Math. USSR Izvestija {\bf 9} (1975) 297.
\bibitem{CecFerGir} S. Cecotti, S. Ferrara and L. Girardello,
Int. J. Mod. Phys. {\bf A4} (1989) 2457.
\bibitem{Cecotti} S. Cecotti, Commun. Math. Phys. {\bf 124} (1989) 23.
\bibitem{ssss} B.~de~Wit, F.~Vanderseypen and A.~Van~Proeyen,
Nucl. Phys. {\bf B400} (1993) 463.
\bibitem{GuSiTo} M.~G\"unaydin, G.~Sierra and P.K.~Townsend, Phys. Lett.
{\bf B133} (1983) 72;
Nucl.~Phys. {\bf B242} (1984) 244, {\bf B253} (1985) 573.
\bibitem{DWVP}  B. de Wit, P.G. Lauwers, R. Philippe, Su S.-Q.
and A. Van Proeyen, Phys. Lett. {\bf 134B} (1984) 37;\\
B. de Wit and A. Van Proeyen, Nucl. Phys. {\bf
B245} (1984) 89.
\bibitem{dWTNic} B.~de~Wit, A.~Tollst\'en and H.~Nicolai, Nucl.
Phys. {\bf B392} (1993) 3.
\bibitem{special} A.~Strominger, Commun.~Math.~Phys. {\bf 133}
(1990) 163.
\bibitem{brokensi} B. de Wit and A. Van Proeyen, Phys. Lett.
{\bf B293} (1992) 94.
\bibitem{dWVP3}  B.~de~Wit and A.~Van~Proeyen,
Commun.~Math.~Phys. {\bf 149} (1992) 307.
\bibitem{Cortes} V. Cort\'es, {\it Alekseevskian spaces}, preprint
Mathematisches Institut Bonn, february 1994,
Diff. Geometry and Applications, to appear.
\bibitem{jrel}  B. de Wit and A. Van Proeyen,
in `Proceedings of the Journ\'ees Relativistes '93', eds. F. Englert,
M. Henneaux and Ph. Spindel, (World Scientific, 1994), pp. 31--47.
Int. J. Mod. Phys. {\bf D3} (1994) 31, hep-th/9310067.
\bibitem{Seiberg} N. Seiberg, Nucl. Phys. {\bf B303} (1988) 286.
\bibitem{FerStro} S.~Ferrara and A.~Strominger, in {\em Strings
'89}, eds. R.~Arnowitt, R.~Bryan, M.J.~Duff, D.V.~Nanopoulos and
C.N.~Pope (World Scientific, 1989), p.~245.
\bibitem{DixKapLou} L.J. Dixon, V.S. Kaplunovsky and J. Louis,
Nucl. Phys. {\bf B329} (1990) 27.
\bibitem{Cand} P.~Candelas and X.~C.~de~la~Ossa, Nucl. Phys. {\bf B355}
(1991) 455,\\
P.~Candelas, X.~C.~de~la~Ossa, P.~Green and
L.~Parkes, Phys.~Lett. {\bf 258B} (1991) 118; Nucl.~Phys. {\bf
B359} (1991) 21.
\bibitem{CdAF} L. Castellani, R. D' Auria and S. Ferrara, Phys. Lett. {\bf
B241} (1990) 57; Cl.Q. Grav. {\bf 7} (1990) 1767,\\
R.~D'Auria, S.~Ferrara and P.~Fr\'e, Nucl.~Phys. {\bf B359}
(1991) 705.
\bibitem{BEC} E. Cremmer, C. Kounnas, A. Van Proeyen, J.P. Derendinger, S.
Ferrara, B. de Wit and L. Girardello, Nucl. Phys. {\bf B250} (1985) 385.
\bibitem{f0art} A. Ceresole, R. D'Auria, S. Ferrara and A. Van
Proeyen, Nucl. Phys. {\bf B444} (1995) 92, hep-th/9502072.
\bibitem{CDFLL} A. Ceresole, R. D'Auria, S. Ferrara, W. Lerche
and J. Louis,  Int. J. Mod. Phys. {\bf A8} (1993) 79, hep-th/9204035.
\bibitem{modssym} A. Ceresole, R. D'Auria and S. Ferrara, Phys. Lett.
{\bf B339} (1994) 71, hep-th/9408036.
\bibitem{CremVP} E. Cremmer and A. Van Proeyen, Class. Quantum Grav. {\bf 2}
(1985) 445.
\bibitem{christoi} C.M. Hull and A. Van Proeyen, Phys. Lett. {\bf
B351} (1995) 188, hep-th/9503022.
\bibitem{Sabi} S. Ferrara and S. Sabharwal, Nucl. Phys. {\bf B332} (1990)
317.
\bibitem{dWVP2} B.~de~Wit and A.~Van~Proeyen, Phys.~Lett. {\bf
B252} (1990) 221.
\bibitem{CliffordR}
M.F. Atiyah, R. Bott and A. Shapiro, Topology {\bf 3},
Sup. 1 (1964) 3.
\bibitem{whatskg} B. Craps, F. Roose, W. Troost and A. Van Proeyen,
hep-th/9703082, to be published in Nucl. Phys. {\bf B}.
\end{thebibliography}
\end{document}